\documentclass[onecolumn,showpacs,superscriptaddress,nofootinbib]{revtex4}
\usepackage{graphicx}  
\usepackage{amssymb}   
\usepackage{bm} 
\usepackage{dcolumn}
\usepackage{color}
\usepackage{mathrsfs}
\usepackage{amsfonts}
\usepackage{varioref}
\usepackage{amsmath}
\RequirePackage[colorlinks,citecolor=blue,urlcolor=magenta,linkcolor=blue]{hyperref}
\def\EH{Einstein-Hilbert }
\def\LL{Lanczos-Lovelock }
\def\gr{general relativity}
\labelformat{section}{Section #1}
\labelformat{subsection}{Section #1}
\labelformat{subsubsection}{Section #1}
\labelformat{subsubsubsection}{Section #1}
\labelformat{equation}{Eq.~(#1)}
\labelformat{eqnarray}{Eq.~(#1)}
\labelformat{figure}{Fig.~#1}
\labelformat{subfigure}{Fig.~\thefigure#1}
\labelformat{table}{Tab.~#1}
\labelformat{appendix}{Appendix #1}
\begin{document}

\title{Buchdahl compactness limit for a pure Lovelock static fluid star}

\author{Naresh Dadhich}
\email{nkd@iucaa.in}
\affiliation{Center for Theoretical Physics, Jamia Millia Islamia, New Delhi 110 025, India}
\affiliation{Inter-University Center for Astronomy and Astrophysics, Post Bag 4 Pune 411 007, India}

\author{Sumanta Chakraborty}
\email{sumantac.physics@gmail.com}
\affiliation{Inter-University Center for Astronomy and Astrophysics, Post Bag 4 Pune 411 007, India}
\affiliation{Department of Theoretical Physics, Indian Association for the Cultivation of Science, Kolkata 700032, India}
\begin{abstract}
We obtain the Buchdahl compactness limit for a pure Lovelock static fluid star and verify that the limit following from the uniform density Schwarzschild's interior solution, which is universal irrespective of the gravitational theory (Einstein or Lovelock), is true in general. In terms of surface potential $\Phi(r)$, it means at the surface of the star $r=r_{0}$, $\Phi(r_{0}) < 2N(d-N-1)/(d-1)^2$ where $d$, $N$ respectively indicate spacetime dimensions and Lovelock order. For a given $N$, $\Phi(r_{0})$ is maximum for $d=2N+2$ while it is always $4/9$, Buchdahl's limit, for $d=3N+1$. It is also remarkable that for $N=1$ Einstein gravity, or for pure Lovelock in $d=3N+1$, Buchdahl's limit is equivalent to the criteria that gravitational field energy exterior to the star is less than half its gravitational mass, having no reference to the interior at all.
\end{abstract}
\maketitle
\section{Introduction}

It is an important question for stellar structure that how compact a static fluid star in equilibrium could be? Buchdahl obtained \cite{Buchdahl:1959zz} the compactness limit by requiring (a) density to be monotonically decreasing outwards, i.e., $d\rho/dr\leq0$ and (b) interior solution of the star is matched to the vacuum Schwarzschild exterior. Then the limit is given by
\begin{equation}\label{Eq12}
r_{0}>\frac{9}{8}(2M);\qquad \Phi(r_{0})\equiv \frac{M}{r_{0}}<\frac{4}{9}~,
\end{equation}
where $M$ and $r_0$ are respectively gravitational mass and radius of the star and $\Phi(r_{0})$ corresponds to gravitational potential at the surface of the star \footnote{Throughout in the literature in some places it is given with inclusion of equality which is not entirely correct because equality corresponds to pressure becoming infinite at the center.}. Clearly, as expected the boundary radius of the star is always greater than the black hole horizon.

Buchdahl compactness limit has attracted a lot of attention in literature and it has been considered in various contexts; inclusion of charge and $\Lambda$ \cite{Mak:2001gg,harko-mak,Andreasson:2012dj,Stuchlik:2008xe}, alternative conditions than that of Buchdahl's \cite{Andreasson:2007ck,Karageorgis:2007cy}, stars on the brane \cite{Germani:2001du,Garcia-Aspeitia:2014pna}, in modified gravity theories \cite{Goswami:2015dma} and in higher dimensions \cite{PoncedeLeon:2000pj,Zarro:2009gd} as well. More importantly, it could be successively pinned down by the assumption of dominant energy condition \cite{Barraco:2002ds} and further requiring sound velocity being subluminal \cite{Fujisawa:2015nda}. Recently it has been studied \cite{Wright:2015yda} for Einstein-Gauss-Bonnet gravity in five dimensions and it is shown that the limit for positive GB coupling depends upon stellar structure, i.e., on the central density.

On the other hand, it has been argued in recent years \cite{Dadhich:2015lra} that pure Lovelock theories with only one $N$th order term in the gravitational action has a number of very interesting and attractive properties. They include --- (a) gravity being kinematic, i.e., there exists no non-trivial vacuum solution (this is due to the fact that Lovelock-Riemann is given entirely in terms of Lovelock-Ricci) in all critical odd $d=2N+1$ dimensions \footnote{This however does not mean that the standard Riemann curvature is zero, rather it is due to a solid angle deficit giving rise to stresses which accord to those of a global monopole. For Gauss-Bonnet gravity it relates to the well known result that vacuum in pure Gauss-Bonnet gravity is highly degenerate \cite{Charmousis:2008ce}. } \cite{Dadhich:2012cv,Camanho:2015hea}; (b) thermodynamic features of \gr\ uphold in Lovelock gravity as well \cite{Chakraborty:2015wma,Chakraborty:2014rga,Chakraborty:2015hna,Chakraborty:2014joa}; (c) unlike Einstein theory 
bound 
orbits around a static source exist in all dimensions $\geq2N+2$ \cite {Dadhich:2013moa}; (d) equipartition of gravitational and non-gravitational energy defines horizon for pure Lovelock black holes \cite{Chakraborty:2015kva,Dadhich:1997ze} and (e) features of geometry quantization and Hawking radiation advertise in favour of pure Lovelock theories \cite{Chakraborty:2016fye,Lochan:2015bha}. Moreover  pure Lovelock static black hole solution with $\Lambda$ asymptotically approximates to the corresponding $d$-dimensional Schwarzschild-de Sitter solution even though the field equations are free of the Einstein term.

With all this, it makes a strong case for pure Lovelock field equations to be the proper equations in higher spacetime dimensions, with $\textrm{dimension}\geq 2N+1$ \cite{Dadhich:2012pd}. This is of course the case only in the context of the classical equations of gravitation in higher dimensions. However this would not be consistent with inclusion of higher order curvature terms --- Lovelock polynomial, as high energy corrections to the Einstein-Hilbert action. In Einstein-Lovelock theory, one can recover the Einstein gravity by switching off the Lovelock part. This is not the case for pure Lovelock theories except the case that static black hole solution asymptotically agrees with the Schwarzschild-de Sitter solution in Einstein gravity.

We shall begin by recapitulating Lovelock gravity and the essential results we require for later parts of this work and then obtain Buchdahl's inequality for both a uniform and non-uniform density star in pure Lovelock gravity. It turns out that when one tries to generalize the Buchdahl's limit to Lovelock gravity in contrast to \gr\ one obtains an additional contribution to the master equation proportional to $(N-1)$. Even though presence of such a term leads to additional difficulties in the derivation of Buchdahl's limit, one can still derive it by demanding a few additional physically motivated requirements on the matter sector. In particular the following conditions will be used in the derivation: (a) Pressure at the center ($p_{c}$) must be finite and positive and in addition it should be less than the central density $\rho _{c}$, i.e., $p_c<\rho_c$; (b) The energy density $\rho$, average density $\bar{\rho}$ and the pressure $p$ shall be decreasing outwards and the rate of change of
pressure ($dp/dr$) should be less than the rate of change of density ($d\rho/dr$), which essentially requires the speed of sound to be less than unity. It is clear that any physically reasonable fluid star will satisfy these requirements. Further the matching conditions between the interior solution with the corresponding exterior vacuum solution will be used extensively. Interestingly it turns out that density profile with both $d\rho/dr=0$ and $d\rho/dr<0$ give the same limit. The one representing constant density leads to an infinite sound velocity and thus implies the limiting degree of compactness. This is because the velocity of sound increases as star gets more and more compact. It is followed in a later section by consideration of gravitational field energy exterior to star. It is remarkable that the Buchdahl limit results when gravitational field energy is less than half of star's gravitational mass for $N=1$, viz. Einstein gravity and for $d=3N+1$, pure Lovelock gravity. We conclude with a 
discussion on our results.
\section{Lovelock gravity: A Brief Revisit}

It is of common belief that at high energy the \EH action is inadequate to describe gravity, higher curvature terms would be indispensable at such scales. However in principle, there can be a large number of choices for such higher curvature terms respecting the diffeomorphism invariance of the action. But most of these higher curvature gravity theories have field equations containing more than second order derivatives of the metric, bringing in ghosts (in other words instabilities) to the theory. Study of such higher curvature theories without any instability has uniquely led us to Lovelock theories of gravity 
\cite{Padmanabhan:2013xyr, Dadhich:2015lra}. The Lovelock polynomial action is given by the Lagrangian
\begin{equation}\label{Eq01}
\mathcal{L} = \sum ^{N}_{n=0} \alpha_{n} \mathcal{R}^{(n)}~,
\end{equation}
where $\alpha _{n}$ are arbitrary coupling constants and are unconstrained a priori. The $N$th order Lovelock Lagrangian $\mathcal{R}^{(N)}$ appearing in the above expression has the following structure,
\begin{equation}
\mathcal{R}^{(N)} = \frac{1}{2^N} \delta^{a_1 b_1 \cdots a_N b_N}_{c_1 d_1 \cdots c_N d_N}\left\lbrace \Pi^N_{r=1} R^{c_r d_r}_{a_r b_r}\right\rbrace~,
\end{equation}
and $R^{ab}_{cd}$ is the standard Riemann tensor (however the same expression can also be written in terms of an analogue of Riemann tensor $^{(n)}\mathcal{R}_{abcd}$ for Lovelock gravity, see for example \cite{Dadhich:2008df}). Further, $ \delta^{a_1 b_1 \cdots a_N b_N}_{c_1 d_1 \cdots c_N d_N} = \frac{1}{N!} \delta^{a_1}_{\left[c_1\right.} \delta^{b_1}_{d_1} \cdots \delta^{a_N}_{c_N} \delta^{b_N}_{\left.d_N \right]}$ is the required determinant tensor, antisymmetric in all the indices. For completeness, let us work out the first three terms in the Lovelock expansion. The $N=0$ case is simple, it just represents the cosmological constant $\Lambda$, while the $N=1$ case yields the standard \EH Lagrangian $R$, i.e., the Ricci scalar. While the situation with $N=2$ involves a particular, quadratic combination of the Riemann tensor and its contracted parts, known as the Gauss-Bonnet Lagrangian, which reads
\begin{equation}
L_{\rm GB}=\mathcal{R}^{(2)}=\frac{1}{4}\delta ^{a_{1}b_{1}a_{2}b_{2}}_{c_{1}d_{1}c_{2}d_{2}}R^{c_{1}d_{1}}_{a_{1}b_{1}}R^{c_{2}d_{2}}_{a_{2}b_{2}}
=R^{2}-4R_{ab}R^{ab}+R_{abcd}R^{abcd}
\end{equation}
Following this prescription one can obtain the higher order terms as well, however they will be more complex and shall depend on higher powers of curvature.

The field equations for any theory can be obtained by varying the corresponding action functional with respect to the dynamical variable. On variation of the Lovelock Lagrangian as presented in \ref{Eq01} including the matter Lagrangian with respect to the metric as the dynamical variable, one obtains the field equations as \cite{Dadhich:2008df}
\begin{equation}
\sum ^N_{n=0} \alpha_n \mathcal{G}^{(n)}_{cd} = \sum ^N_{n=0} \alpha_n \left(n\mathcal{R}^{(n)}_{cd}-\frac{1}{2}\mathcal{R}^{(n)}g_{cd}\right) =8\pi T_{cd}~,
\end{equation}
where $ \mathcal{R}^{(n)}_{cd} = g^{pq}\mathcal{R}^{(n)}_{cpdq}, \mathcal{R}^{(n)}=g^{cd}\mathcal{R}^{(n)}_{cd}$, and $T_{cd}$ is the matter energy momentum tensor. This is the gravitational field equations in Lovelock gravity, which is trivial for $N=0$, lead to Einstein's equations for $N=1$ and to Gauss-Bonnet field equations for $N=2$, and so on. In particular for the Einstein-Gauss-Bonnet gravity, the field equations sketched above take the following form,
\begin{equation}
G^{a}_{b}+\alpha_{\rm GB} H^{a}_{b}=8\pi T^{a}_{b}~,
\end{equation}
where, $\alpha_{\rm GB}$ stands for the Gauss-Bonnet coupling parameter and $\mathcal{G}^{(2)}_{ab}=H_{ab}$, having the following expression,
\begin{equation}
H_{ab}=2\left(RR_{ab}-2R_{ac}R^{c}_{b}-2R^{cd}R_{acbd}+R^{~pqr}_{a}R_{bpqr}\right)-\frac{1}{2}g_{ab}\mathcal{R}^{(2)}~.
\end{equation}
In the following, we shall specialize to pure Lovelock theories with $N$ fixed and shall assume perfect fluid matter distribution. Note that in the case of Einstein-Lovelock theories of order $N$, there are total $N$ arbitrary constants, among which only one appears in the coupling with matter and hence gets determined by the experiment. However the other $(N-1)$ constants remain completely arbitrary. On the other hand, in the case of pure Lovelock theories there is a single coupling constant and this is what appears in the gravitational potential and acts as a substitute for the Newton's constant in higher dimensions. The field equations should therefore have either only one coupling constant which could be fixed by experiment by measuring the strength of gravitational interaction \cite{Dadhich:2012pd} or all other couplings are given in terms of the absolute vacuum $\Lambda$ as was the case for dimensionally continued black hole solutions \cite{Banados:1992wn}. 
\section{Buchdahl's Compactness Limit}

We shall in the following obtain Buchdahl's compactness limit not only for Einstein but for pure Lovelock static star in hydrostatic equilibrium. As a warm up we will first discuss the case of a uniform density star, making way for variable density star in the next section. Finally we will show that similar limits can be obtained from consideration of gravitational field energy as well.

At the outset, let us note that since gravity is kinematic in critical odd $d=2N+1$ dimensions \cite{Dadhich:2012cv}, there can occur no bound fluid distribution for a star interior as we shall just see. The spacetime being static and spherically symmetric, the metric describing it can be written in the following form,
\begin{equation}
ds^{2}=-e^{\nu}dt^{2}+e^{\lambda}dr^{2}+r^{2}d\Omega_{d-2}^{2}~,
\end{equation}
where, $e^{\nu}$ and $e^{\lambda}$ are dependent on the radial coordinate $r$ and $d\Omega _{d-2}^{2}$ represents the line element on the surface of a $(d-2)$ dimensional sphere. In pure Lovelock theories, for the above metric ansatz one can obtain analytical expressions for the left hand side (i.e., geometric part) of the field equations. On the other hand, the right hand side of the field equations shall contain information about the matter. Since the perfect fluid acting as the source for matter, is isotropic, one obtains, the only nontrivial components of the matter energy-momentum tensor to be, $T^{0}_{0}=-\rho$ and $T^{r}_{r}=p$ respectively. Hence, the corresponding field equations are (see \ref{App_A} for a detailed derivation),
\begin{equation}\label{Eq02}
8\pi \rho(r)=\frac{\left(1-e^{-\lambda}\right)^{N-1}}{2^{N-1}r^{2N}}\left[rN\lambda'e^{-\lambda}+(d-2N-1)\left(1-e^{-\lambda}\right)\right]~,
\end{equation}
and
\begin{equation}\label{Eq03}
8\pi p(r)=\frac{\left(1-e^{-\lambda}\right)^{N-1}}{2^{N-1}r^{2N}}\left[rN\nu'e^{-\lambda}-(d-2N-1)\left(1-e^{-\lambda}\right)\right]~.
\end{equation}
where the corresponding coupling parameter $\alpha _{N}$ has been chosen accordingly. On the other hand, from conservation of the matter energy momentum tensor, i.e., $\nabla _{\mu}T^{\mu}_{\nu}=0$, we obtain
\begin{equation}\label{Eq04}
2p'=-\nu '\left(\rho+p\right)~.
\end{equation}
Note that in all the previous expressions `prime' denotes derivative with respect to the radial coordinate $r$. It is evident that, one can immediately integrate \ref{Eq02}, which results into,
\begin{equation}\label{Eq05}
e^{-\lambda}=1-\left(\frac{2^{N}m(r)}{r^{d-2N-1}}\right)^{1/N};\qquad m(r)=4\pi \int ^{r} \rho (r)r^{d-2}dr~.
\end{equation}
While in order to solve for $\nu$, we will follow the approach in \cite{Wald,gravitation}. Using the expressions for energy density and pressure from \ref{Eq02} and \ref{Eq03}, one can substitute them in the continuity equation, i.e., \ref{Eq04}, leading to,
\begin{eqnarray}
2r\nu''+r\nu'^{2}&-&2\nu'-4(N-1)\nu'-r\nu'\lambda'+2r\nu'\lambda'\left[1+\frac{Ne^{-\lambda}-1}{1-e^{-\lambda}}\right]~,
\nonumber
\\
&=&\frac{2(d-2N-1)e^{\lambda}}{N}\left[N\lambda'e^{-\lambda}-\frac{2N}{r}\left(1-e^{-\lambda}\right)\right]~.
\end{eqnarray}
This is the pressure isotropy equation. One can use the following two identities,
\begin{eqnarray}
\frac{d}{dr}\left[\frac{1}{r}e^{-\lambda/2}\frac{de^{\nu/2}}{dr}\right]
&=&\frac{e^{(\nu-\lambda)/2}}{4r^{2}}\left[2r\nu''+r\nu'^{2}-2\nu'-r\nu'\lambda'\right]~,
\\
\frac{d}{dr}\left[\frac{1-e^{-\lambda}}{2r^{2}}\right]&=&\frac{e^{-\lambda}}{2r^{3}}\left[r\lambda'-2\left(e^{\lambda}-1\right)\right]~,
\end{eqnarray}
for casting the above equation in the following form
\begin{eqnarray}\label{Eq06}
e^{-(\nu+\lambda)/2}&\times&\frac{d}{dr}\left[\frac{1}{r}e^{-\lambda/2}\frac{de^{\nu/2}}{dr}\right]
\nonumber
\\
&=&\left[(d-2N-1)-(N-1)\frac{r\nu'}{e^{\lambda}-1}\right]\frac{d}{dr}\left[\frac{1-e^{-\lambda}}{2r^{2}}\right]
\nonumber
\\
&=&\left[(d-2N-1)-(N-1)\frac{r\nu'}{e^{\lambda}-1}\right]\frac{\bar{\rho}^{\frac{1-N}{N}}}{N}\frac{d\bar{\rho}}{dr}~.
\end{eqnarray}
In order to arrive at the last line, we have used \ref{Eq05}, to immediately obtain, $(1-e^{-\lambda})/2r^{2}=(m(r)/r^{d-1})^{1/N}$. From the definition for $m(r)$ in \ref{Eq05}, it is evident that one can define an average density as, $\bar{\rho}=m(r)/r^{d-1}$, and hence Eq. (\ref{Eq06}) follows. We will now discuss the two situations, first with uniform density and then with the variable one.
\subsection{From uniform density}

The uniform density sphere is always given by Schwarzschild interior solution irrespective of gravity being described by Einstein or Lovelock theory \cite{Dadhich:2010qh} and it reads as follows in a $d$ dimensional spacetime:
\begin{equation}\label{Eq07}
e^{\nu/2}=A+Be^{-\lambda/2};\qquad  e^{-\lambda}=1-\mu r^{2}~.
\end{equation}
Here $A$ and $B$ are constants to be determined by matching the interior solution to the exterior vacuum solution, where star boundary is defined by $p=0$, and constant $\mu$ refers to uniform density of the matter distribution.

Since density is constant, \ref{Eq05} yields,
\begin{equation}
e^{-\lambda}=1-\mu r^2; \qquad \mu=2\left(\frac{4\pi\rho}{(d-1)}\right)^{1/N}~,
\end{equation}
which matches exactly with \ref{Eq07}. It is clear from expression for radial pressure, i.e., \ref{Eq03} that in critical odd $d=2N+1$ dimensions, there cannot occur bound distribution because at the boundary $p=0$, which leads to $\nu'=0$. However matching to vacuum exterior would demand $\nu'\neq0$. Thus in critical odd dimension there can exist no static star interior \cite{Dadhich:2015rea}. Note that this is true, irrespective of whether $\rho$ is constant or not.

One glance at Eq. (\ref{Eq06}) reveals that for constant density the right hand side of Eq. (\ref{Eq06}) identically vanishes, thus the left hand side trivially integrates to give $e^{\nu}$ as presented in \ref{Eq07}. This is the well known Schwarzschild interior solution, which is unique for distributions having constant density, irrespective of gravitational theory (Einstein or Lovelock \footnote{Though here we have considered pure Lovelock equations, it is remarkable that even for Einstein-Lovelock with sum over $N$, the solution remains the same - Schwarzschild's \cite{Dadhich:2010qh}.}). As already emphasized, there can exist no bound distributions in the critical odd $d=2N+1$ dimension, we shall henceforth consider $d\geq2N+2$.

As argued earlier that uniform density marks the limiting degree of compactness, hence the Schwarzschild solution describing uniform density star should give the compactness limit, and it does indeed do that \cite{Wald}. The interior solution of the star (as in \ref{Eq07}) has to be matched with the outside vacuum solution, which for pure Lovelock theory would be given by \cite{Dadhich:2012ma}
\begin{equation}
e^{\nu}=1-\frac{2M^{1/N}}{r^{\alpha}}=e^{-\lambda};\qquad \alpha=\frac{(d-2N-1)}{N}~.
\end{equation}
Note that ADM mass in this context is $M=(M^{1/N})^N$ \cite{Kastor:2008xb}. Now continuity of metric and $\nu'$, which is equivalent to $p=0$ at the boundary of the star, say at $r=r_{0}$, determines the constants $A$ and $B$ as
\begin{equation}
A=\frac{d-1}{2N}\sqrt{1-2\Phi};\qquad  \Phi=M^{1/N}/r_{0}^{\alpha}~,
\end{equation}
and
\begin{equation}
B=-\frac{d-2N-1}{2N}~,
\end{equation}
where, we have used the result that, $\mu=(M^{1/N}/r_{0}^{\alpha+2})$. In the interior region one can use \ref{Eq03}, in order to obtain the condition for pressure to be finite and positive as, $rN\nu'>(d-2N-1)(e^{\lambda}-1)$. Expressing this inequality using the solutions for the metric elements as presented in \ref{Eq07} and evaluating at the center we arrive at, $1>A-|B|>0$. The condition $A-|B|>0$, leads to,
\begin{equation}\label{Eq11}
(d-1)\sqrt{1-2\Phi}>(d-2N-1)~.
\end{equation}
While the condition, $A-|B|<1$, simplifies to, $(d-1)(\sqrt{1-2\Phi}-1)<0$, which is identically satisfied, since $\Phi>0$. Finally, \ref{Eq11} readily leads to the required Buchdahl limit as
\begin{equation}\label{Eq13}
r_{0}^{\alpha}>\frac{(d-1)^2}{2N(d-N-1)}M^{1/N};\qquad \Phi<\frac{2N(d-N-1)}{(d-1)^2}~.
\end{equation}
Note that it assumes the familiar form $\Phi<4/9$ as in \ref{Eq12} for the dimensional spectrum $d=3N+1$ including of course $N=1, d=4$. However, the derivation presented above hinges on a very crucial assumption of $\rho=\textrm{constant}$, which seems physically unrealistic. Thus we will now take up the case for variable density star and show that the same inequality follows.
\subsection{From non uniform density}

In this section we will derive Buchdahl's inequality for a general fluid, with its density and pressure decreasing outwards. In this general situation, it is not possible to integrate Eq. (\ref{Eq06}) and obtain $e^{\nu}$ in a closed form. In the absence of exact solution we will consider the technique used by \cite{Wald,gravitation}. This method essentially requires the right hand side of Eq. (\ref{Eq06}) to be negative, which follows from $d\bar{\rho}/dr<0$ provided the coefficient multiplying it is positive. However for pure Lovelock theories, with $N>1$ there is no guarantee that this coefficient would be positive. However for Einstein gravity, $N=1$, it is obviously so. Thus one needs to either show that despite the presence of a negative factor that coefficient is positive and hence the same argument should work on Lovelock gravity as well, or it will provide some constraints on the possible geometry which would satisfy Buchdahl's limit. We will return to this issue later on, but for the
present, we will assume it to be positive and proceed. In that case, we have the following inequality,
\begin{equation}\label{Eq10}
e^{-(\nu+\lambda)/2}\frac{d}{dr}\left[\frac{1}{r}e^{-\lambda/2}\frac{de^{\nu/2}}{dr}\right]\leq 0~.
\end{equation}
Integrating this inequality inwards, i.e., from radius $r_{0}$ to radius $r$. we obtain,
\begin{equation}
\frac{1}{r}e^{-\lambda/2}\frac{de^{\nu/2}}{dr}\geq \frac{1}{r_{0}}\left[e^{-\lambda/2}
\frac{d}{dr}e^{\nu /2}\right]_{r=r_{0}}
=\frac{\alpha M^{1/N}}{r_{0}^{\alpha+2}}~,
\end{equation}
where continuity of the metric and its first derivative has been assumed. Multiplying the equation by $re^{\lambda /2}$ and again integrating inwards from $r=r_{0}$ to $r=0$, we obtain,
\begin{equation}\label{Eq08}
e^{\nu/2}(r=0)\leq \sqrt{1-\left(\frac{2M^{1/N}}{r_{0}^{\alpha}}\right)}
-\frac{\alpha M^{1/N}}{r_{0}^{\alpha+2}}\int _{0}^{r_{0}}dr~r\left\lbrace1-\left(\frac{2m(r)^{1/N}}{r^{\alpha}}\right)\right\rbrace ^{-1/2}~.
\end{equation}
Since, we have $d\bar{\rho}/dr<0$, it immediately follows that, $m(r)$ must be larger than it would have been for a uniform density star, such that,
\begin{equation}
m(r)\geq \frac{Mr^{d-1}}{r_{0}^{d-1}}~.
\end{equation}
With this condition \ref{Eq08} holds more strongly and hence we obtain,
\begin{eqnarray}
e^{\nu/2}(r=0)&\leq& \sqrt{1-\left(\frac{2M^{1/N}}{r_{0}^{\alpha}}\right)}
-\frac{\alpha M^{1/N}}{r_{0}^{\alpha+2}}\int _{0}^{r_{0}}dr~r\left\lbrace 1-\frac{2M^{1/N}}{r_{0}^{(d-1)/N}}r^{2}\right\rbrace ^{-1/2}
\nonumber
\\
&=&\left(1+\frac{\alpha}{2}\right)\sqrt{1-\left(\frac{2M^{1/N}}{r_{0}^{\alpha}}\right)}
-\frac{\alpha}{2}~.
\end{eqnarray}
For pressure to be finite and positive at the center, one must have $e^{\nu/2}(r=0)>0$ \footnote{Near, $r=0$, $(1-e^{-\lambda})$ must scale as $r^{2}$ to ensure that pressure at the center remains finite. Thus, $e^{\nu/2}(r_{0})\sim \sqrt{1-2\Phi}>0$, which is a strict inequality.} and hence the above inequality as applied to the last line would imply
\begin{equation}
\sqrt{1-\left(\frac{2M^{1/N}}{r_{0}^{\alpha}}\right)}>\frac{\alpha}{2+\alpha}~,
\end{equation}
which simplifies to
\begin{equation}\label{Eq09}
M^{1/N}<\frac{2N(d-N-1)}{(d-1)^{2}}r_{0}^{\frac{d-2N-1}{N}}~.
\end{equation}
This is the same inequality derived for constant density as given in \ref{Eq13} and it reduces to the familiar form $M < (4/9)r_{0}$ for Einstein gravity in four spacetime dimension, i.e., for $N=1$ and $d=4$.

It is remarkable that for the dimensional spectrum, $d=3N+1$, which gives $1/r$ potential \cite{Chakraborty:2016qbw} for pure Lovelock vacuum, $M^{1/N}<(4/9)r_{0}$ always. Also note that $d\geq(3N+1)$ defines the stability threshold for pure Lovelock star. For a given $N$, the star is most compact in the critical $d=2N+2$ dimensions, but it is unstable \cite{Gannouji:2013eka}. On the other hand, $M^{1/N}r_{0}^{-\alpha}\to 1/2$, the black hole limit, as $N\to\infty$. Thus the compactness limit is universal for the stability threshold dimension, $d=3N+1$. For instance, for $N=2$ Gauss Bonnet gravity, we shall have $\sqrt{M}/r<4/9$ for $d=7$. Hence we have derived Buchdahl's inequality for a star in pure Lovelock gravity and it turns out that for the dimensional spectrum $d=3N+1$, it is always $\Phi<4/9$.

We will now argue for the coefficient of $d\bar{\rho}/dr$ to be positive, in the context of pure Lovelock gravity. Let us first see what happens at the surface $r=r_{0}$. Since at the boundary of the star pressure must vanish, from \ref{Eq03} we immediately obtain, $r\nu'/(e^{\lambda}-1)=(d-2N-1)/N$. As a consequence one obtains,
\begin{equation}
(d-2N-1)-(N-1)\frac{r\nu'}{e^{\lambda}-1}=\frac{(d-2N-1)}{N}>0~.
\end{equation}
Hence it is positive at the boundary.

The only remaining bit corresponds to showing the monotonicity of the function $r\nu'/(e^{\lambda}-1)$, which can be achieved in a straightforward manner by taking radial derivative of \ref{Eq03}, which leads to,
\begin{equation}
\frac{2Nm(r)}{r^{d-1}}\frac{d}{dr}\left(\frac{r\nu'}{e^{\lambda}-1}\right)=8\pi \frac{dp}{dr}-\frac{8\pi p(r)r^{d-1}}{m(r)}\frac{d\bar{\rho}}{dr}=8\pi p\frac{d}{dr}{\ln \beta}
\end{equation}
where $\beta =p/\bar\rho$ and recall that $\bar\rho = m(r)/r^{d-1}$. Further, we assume that the gradient of pressure ($dp/dr$) is smaller compared to the gradient of density $d\rho/dr$, implying the ratio $p/\bar{\rho}$ to decrease as one moves outwards, signaling $d\beta/dr<0$. Also note that $0\leq p \leq \bar{\rho}$ is one of the reasonable conditions that ensure sound velocity to be less than that of light. This suggests that for physically realistic fluid interior of a star, $\beta$ starts with some finite value at the center then decreases with pressure, ultimately vanishing at the boundary along with pressure. Using these results in the above equation one immediately obtains,
\begin{equation}
\frac{d}{dr}\left(\frac{r\nu'}{e^{\lambda}-1}\right)<0
\end{equation}
Hence the ratio $r\nu'/(e^{\lambda}-1)$ is a monotonic function and decreases outwards. In the above we have already shown positivity of the coefficient of $d\bar{\rho}/dr$ at the boundary, while if one can now prove positivity also at the center, then the monotonicity will ensure that it remains positive throughout the star as well.

For that purpose, one can consider the ratio of pressure and density at the central region. Let the central pressure be given by $p(0)=p_{c}$ and the corresponding density being $\rho _{c}$. Then we obtain, near $r=0$, $1-e^{-\lambda}=2r^{2}(4\pi \rho _{c}/(d-1))^{1/N}$. Substitution of which in \ref{Eq03}, i.e., the expression for pressure, we obtain,
\begin{equation}
\frac{p_{c}}{\rho _{c}}=\frac{1}{(d-1)}\left[N\frac{r\nu'}{e^{\lambda}-1}-(d-2N-1)\right]~.
\end{equation}
Now $p_{c}/\rho _{c}<1$ requires $r\nu'/(e^{\lambda}-1)<(d-2N-1)/(N-1)$. This completes the proof of the coefficient of $d\bar{\rho}/dr$ being positive and hence the Buchdahl inequality follows from the general case of non-uniform density. Note that apart from density and pressure decreasing outwards, we also require $p<\rho$.
\subsection{From gravitational field energy}

In this section, we will present yet another derivation for Buchdahl's inequality, albeit from a completely different perspective. It is well known that there exists no covariant definition for gravitational field energy in general relativity. However there are a few very useful and insightful prescriptions for some situations of physical interest, for example Komar integral for static vacuum spacetime giving conserved mass for a black hole \cite{Komar:1958wp,Kulkarni:1988CQG} and similarly Brown-York quasi-local energy \cite{Brown:1992br} giving measure of total energy including both matter and gravity contained inside a radius $r$. From the latter it is straightforward to find gravitational field energy lying outside a static star, which we shall employ.

In the Brown-York prescription, it is envisioned that a spacetime region is bounded in a $3$-cylindrical timelike surface which is bounded by a $2$-surface at the two ends. Then Brown-York quasi-local energy is defined by \cite{Brown:1992br},
\begin{equation}
E_{\rm BY}= \frac{1}{8\pi} \int{d^2x\sqrt{q}(k-k_0)}~,
\end{equation}
where $k$ and $q_{ab}$ are respectively, trace of extrinsic curvature and metric on $2$-surface. Here $k_0$ refers to some reference spacetime, for instance it would naturally be Minkowski flat for asymptotically flat spacetimes. This is the measure of total energy contained inside some radius $r$, and for a static body described by Schwarzschild exterior solution, it is given by
\begin{equation}
E_{\rm BY}=r\left(1-\sqrt{1-2M/r}\right)~.
\end{equation}
Gravitational field energy exterior to a sphere of radial extent $r$ would simply be given by subtracting mass $M$ we started with from it, and hence we can write
\begin{equation}
E_{\rm GF}=E_{\rm BY}-M=r\left(1-\sqrt{1-2M/r}\right)-M~.
\end{equation}
This was very imaginatively employed in defining black hole horizon \cite{Dadhich:1997ze}, where it was argued that timelike particles feel potential gradient produced by gravitating mass while photons can only respond to space curvature produced by gravitational field energy \cite{Dadhich:1997ku,Dadhich:2012pda}. The most remarkable thing that happens is that $E_{GF}<M/2$ defines Buchdahl compactness limit for a star interior; that is,
\begin{equation}
E_{\rm GF}=E_{\rm BY}-M=r\left(1-\sqrt{1-2M/r}\right)-M<\frac{M}{2}~,
\end{equation}
yielding the limiting radius $r_{0}$ as
\begin{equation}
r_{0}>\frac{9}{8}(2M);\qquad \Phi(r_{0})=\frac{M}{r_{0}}<\frac{4}{9}~.
\end{equation}
This however has no reference to star interior. It is perhaps because binding energy in concept and in measure is similar to gravitational field energy. It is however remarkable that Schwarzschild metric in exterior also governs, how compact a star interior could be? This definition also works in higher dimensions, where we write
\begin{equation}
E_{\rm GF}=E_{\rm BY}-M=r^{d-3}\left(1-\sqrt{1-2M/r^{d-3}}\right)-M<\frac{M}{2}~,
\end{equation}
leading to,
\begin{equation}
r_{0}^{d-3}>\frac{9}{8}(2M);\qquad \Phi(r_0)=\frac{M}{r_{0}^{d-3}}<\frac{4}{9}~.
\end{equation}
Hence for Einstein gravity, $E_{\rm GF}<M/2$ always defines the Buchdahl compactness limit in all dimensions $d\geq4$. The question arises, does this work for Lovelock gravity as well?

Very recently we have also extended the above Brown-York energy and hence the gravitational field energy consideration to Lovelock gravity \cite{Gravanis:2010zz,Chakraborty:2015kva}. As stated in the introduction, pure Lovelock gravity has several remarkable distinguishing features, and to that impressive list we add one more and perhaps a very interesting one. In this case gravitational field energy is given by
\begin{equation}
E_{\rm GF}=E_{\rm BY}-M^{1/N}=r^{\alpha}\left(1-\sqrt{1-\frac{2M^{1/N}}{r^{\alpha}}}\right)-M^{1/N}~.
\end{equation}
Clearly when it is equal to $M^{1/N}$, black hole horizon $r^{\alpha}=2M^{1/N}$ is defined and when it is less than $M/2$, we have
\begin{equation}
r^{\alpha}>\frac{9}{8}(2M^{1/N})~,
\end{equation}
as obtained earlier. This however does not agree with the compactness limit as presented in \ref{Eq09}, except for $d=3N+1$, when it reads as $\Phi(r_{0})=M^{1/N}/r_{0}<4/9$. Thus $E_{\rm GF}<M^{1/N}/2$ defines the compactness limit either for Einstein gravity in $d\geq4$ or for pure Lovelock gravity only in $d=3N+1$.

As an aside, we would like to mention that for construction of maximum mass star in a realistic situation obeying dominant energy condition \cite{Barraco:2002ds} and sound velocity being subluminal \cite{Fujisawa:2015nda}, the actual limit is less than Buchdahl limit. It is $M/r<3/8<4/9$ for the former while for the latter it is $M/r<0.3636<3/8$. Buchdahl's limit is however an all covering limit representing an asymptotic state. It is natural to expect that radius of a star should always be greater than unstable photon orbit radius $3M$, which indeed is the case, save for a narrow parameter window, as numerically shown in \cite{Fujisawa:2015nda}.
\section{Discussion}

It is well known that Buchdahl compactness limit which was obtained by the condition $d\rho/dr\leq0$ could as well be obtained from the Schwarzschild interior solution describing uniform density star \cite{Buchdahl:1959zz,Wald,Dadhich:2010qh}. The interesting point is that the inequality in the condition also yields the same limit. This means uniform density, though not entirely physically reasonable, seems to capture the essential aspects of the interior structure of a star. This happens not only for Einstein gravity in $d\geq4$ but also for pure Lovelock gravity, as we have shown.

On the other hand, a more physically realistic situation corresponds to a variable density star such that the density decreases outwards, i.e., $d\rho/dr<0$. For Einstein gravity, in any spacetime dimension, the above condition is sufficient to derive Buchdahl's inequality. It turns out that in the case of pure Lovelock gravity, one needs two additional conditions, namely $p < \rho$ and the pressure decreasing faster compared to the density. These conditions in fact ensures that sound speed is less than that of light. These are therefore very strongly motivated physical conditions. With these conditions we have explicitly derived Buchdahl's inequality for fluid star of non-uniform density in pure Lovelock gravity, which matches exactly with the limit derived for uniform density star. It is however interesting to see that for pure Lovelock theories, existence of Buchdahl's inequality demands the speed of sound to be less than that of light.

It turns out that for a given Lovelock order $N$, the most compact star would exist in $d=2N+2$, while $\Phi(r_{0})\to 1/2$ defines the black hole horizon as $N\to\infty$. This being the absolute cut off bound for compactness. On the other hand, the dimensional spectrum $d=3N+1$ is distinguished for (a) $\Phi(r_{0})=\lbrace M^{1/N}/r\rbrace <4/9$ always, like in the case of \gr, (b) it also defines stability threshold \cite{Chakraborty:2016qbw}.

Alternatively, we have also obtained the compactness limit for Einstein and Lovelock gravity by requiring gravitational field energy being less than half of gravitational mass of the star. However, for pure Lovelock theories it gives the correct limit only for $d=3N+1$. It is however remarkable that measure of gravitational energy in the exterior also determines how compact an interior could be? Why should it happen and more importantly what does the limit $M/2$ physically signify? These are the questions we do not quite understand. It may however be thought that binding energy and gravitational field energy are both similar in concept and perhaps in measure as well. Of course, binding energy is the natural agent to govern compactness \cite{Breu:2016ufb} while gravitational energy should rather govern curvature of space around star \cite{Dadhich:2012pda}. It seems to indicate that both binding and gravitational energy have the same measure. Even then what does this limit physically signify? This is another 
interesting result that points to some deeper connection, and it raises an exciting question that asks for further and serious consideration.
\section*{Acknowledgment}

ND warmly thanks Albert Einstein Institute, Golm for a summer visit that has facilitated this work. He also thanks Rituparna Goswami for discussions and for drawing attention to this problem. Thanks are also due to Xian Camanho for discussion and Luciano Rezzolla for discussion and an engaging dialogue on email as well as for pointing out the references \cite{Barraco:2002ds,Fujisawa:2015nda}, Hiromi Saida for a useful clarification by email exchange and Sourav Bhattacharya for pointing out an important reference. SC thanks CSIR, Government of India for providing a SPM fellowship.
\appendix
\labelformat{section}{Appendix #1}
\labelformat{subsection}{Appendix #1}
\labelformat{subsubsection}{Appendix #1}
\section*{Appendices}

\section{Field equations in pure Lovelock in presence of spherical symmetry} \label{App_A}

The field equations for \LL gravity of order $m$ can be written as
\begin{align}\label{pure_Eq_01}
\mathcal{G}^{a~(m)}_{b}\equiv -\frac{1}{2^{m+1}}\delta ^{ac_{1}d_{1}\ldots c_{m}d_{m}}_{ba_{1}b_{1}\ldots a_{m}b_{m}}R^{a_{1}b_{1}}_{c_{1}d_{1}}\ldots R^{a_{m}b_{m}}_{c_{m}d_{m}}=8\pi \kappa_{(m)} T^{a}_{b}
\end{align}
where, $\kappa _{(m)}=\alpha _{m}^{-1}$. We will use this expression to write down the components of $\mathcal{G}^{a~(m)}_{b}$ in $d$ spacetime dimensions suited for a static and spherically symmetric metric ansatz
\begin{align}\label{pure_ansatz}
ds^{2}=-e^{\nu(r)}dt^{2}+e^{\lambda(r)}dr^{2}+r^{2}d\Omega_{d-2}^{2}
\end{align}
Since we have two unknown functions $\lambda(r)$ and $\nu(r)$ to solve for, evaluation of two components of \ref{pure_Eq_01} would suffice to get an expression for both. We choose these components to be the $(t,t)$ and $(r,r)$ components of the Lovelock field equations, which read,
\begin{align}\label{pure_Eqtt_02}
G^{t~(m)}_{t}&=-\frac{1}{2^{m+1}}\left[4m\left(\delta ^{trA_{2}A_{3}\ldots A_{2m}}_{trB_{2}B_{3}\ldots B_{2m}}\right)R^{rB_{2}}_{rA_{2}}\ldots R^{B_{2m-1}B_{2m}}_{A_{2m-1}A_{2m}}+\left(\delta ^{tA_{1}A_{2}\ldots A_{2m-1}A_{2m}}_{tB_{1}B_{2}\ldots B_{2m-1}B_{2m}}\right)R^{B_{1}B_{2}}_{A_{1}A_{2}}\ldots R^{B_{2m-1}B_{2m}}_{A_{2m-1}A_{2m}}\right]
\\
G^{r~(m)}_{r}&=-\frac{1}{2^{m+1}}\left[4m\left(\delta ^{rtA_{2}A_{3}\ldots A_{2m}}_{rtB_{2}B_{3}\ldots B_{2m}}\right)R^{tB_{2}}_{tA_{2}}\ldots R^{B_{2m-1}B_{2m}}_{A_{2m-1}A_{2m}}+\left(\delta ^{rA_{1}A_{2}\ldots A_{2m-1}A_{2m}}_{rB_{1}B_{2}\ldots B_{2m-1}B_{2m}}\right)R^{B_{1}B_{2}}_{A_{1}A_{2}}\ldots R^{B_{2m-1}B_{2m}}_{A_{2m-1}A_{2m}}\right]
\label{pure_Eqtt_02b}
\end{align}
In order to evaluate the above expressions explicitly we need to have the following components of the Riemann curvature tensor \cite{Chakraborty:2015taq}
\begin{align}
R^{tA}_{tB}=\frac{\nu'e^{-\lambda}}{2r}\delta ^{A}_{B};\qquad R^{rA}_{rB}=\frac{\lambda'e^{-\lambda}}{2r}\delta ^{A}_{B};\qquad R^{AB}_{CD}=\frac{1-e^{-\lambda}}{r^{2}}\delta ^{AB}_{CD}
\end{align}
Note that in the above expressions for Riemann curvature tensors capital Roman indices stand for the $(d-2)$ angular coordinates, a convention we would follow throughout this section. On using these Riemann tensor components, the $(t,t)$ component of $m$th order Lovelock tensor from \ref{pure_Eqtt_02} turns out to yield,
\begin{align}
G^{t}_{t}=-\frac{1}{2^{m+1}}\left[\frac{m2^{m}\lambda'e^{-\lambda}(1-e^{-\lambda})^{m-1}}{r^{2m-1}}\Delta _{(2m-1)}+\frac{(1-e^{-\lambda})^{m}2^{m}}{r^{2m}}\Delta _{(2m)}\right]
\end{align}
where we have defined,
\begin{align}
\Delta _{(k)}=\delta ^{a_{1}a_{2}\ldots a_{k}}_{b_{1}b_{2}\ldots b_{k}}\delta ^{b_{1}}_{a_{1}}\ldots \delta ^{b_{k}}_{a_{k}}
\end{align}
which in $d$ dimension satisfies the recursion relation $\Delta _{(k)}=(d-k-1)\Delta _{(k-1)}$. The derivation essentially hinges on expansion of the full antisymmetric tensor and then contraction with each Kronecker delta separately \cite{Paranjape:2006ca}. Use of this recursion relation repeatedly simplifies the $(t,t)$ component of field equations, which finally reads,
\begin{align}\label{pure_finalEq_03}
\frac{(d-2)!}{(d-2m-1)!}\alpha _{m}\frac{(1-e^{-\lambda})^{m-1}}{2r^{2m}}\Bigg[-mr\lambda'e^{-\lambda}-(d-2m-1)(1-e^{-\lambda})\Bigg]=8\pi T^{t}_{t}
\end{align}
Let us now concentrate on the $(r,r)$ component, which can be derived starting from \ref{pure_Eqtt_02} following the previous root, leading to,
\begin{align}
\frac{(d-2)!}{(d-2m-1)!}\alpha _{m}\frac{(1-e^{-\lambda})^{m-1}}{2r^{2m}}\Bigg[mr\nu'e^{-\lambda}-(d-2m-1)(1-e^{-\lambda})\Bigg]=8\pi T^{r}_{r}
\end{align}
Setting the gravitational coupling $\alpha _{m}$, such that, $\lbrace (d-2)!/(d-2m-1)!\rbrace 2^{m-2}\alpha _{m}=1$, one immediately arrives at \ref{Eq02} and \ref{Eq03} respectively.
\bibliography{Gravity_1_full,Gravity_2_partial,My_References}

\bibliographystyle{./utphys1}
\end{document}